\documentclass[journal]{vgtc}                     

\onlineid{1567}

\vgtccategory{Research}

\title{Data Formulator: AI-powered Concept-driven Visualization Authoring}

\author{%
  \authororcid{Chenglong Wang}{0000-0002-5933-6620}, \authororcid{John Thompson}{0000-0002-3102-4035}, and \authororcid{Bongshin Lee}{0000-0002-4217-627X} 
}

\authorfooter{
  \item
  	Chenglong Wang, John Thompson, and Bongshin Lee are with Microsoft Research.
  	E-mail: \{chenglong.wang, johnthompson, bongshin\}@microsoft.com.
}

\abstract{%
  With most modern visualization tools, authors need to transform their data into tidy formats to create visualizations they want. Because this requires experience with programming or separate data processing tools, data transformation remains a barrier in visualization authoring. To address this challenge, we present a new visualization paradigm, \emph{concept binding}, that separates high-level visualization intents and low-level data transformation steps, leveraging an AI agent. We realize this paradigm in Data Formulator, an interactive visualization authoring tool. With Data Formulator, authors first define \emph{data concepts} they plan to visualize using natural languages or examples, and then bind them to visual channels. Data Formulator then dispatches its AI-agent to automatically transform the input data to surface these concepts and generate desired visualizations.
  When presenting the results (transformed table and output visualizations) from the AI agent, Data Formulator provides feedback to help authors inspect and understand them. A user study with 10 participants shows that participants could learn and use Data Formulator to create visualizations that involve challenging data transformations, and presents interesting future research directions. 
}


\usepackage{tabu}                      
\usepackage{booktabs}                  
\usepackage{lipsum}                    
\usepackage{mwe}                       

\usepackage{mathptmx}                  
\usepackage{stackengine}

\usepackage[frozencache=true,cachedir=minted-cache]{minted}
\setminted{
    fontsize=\footnotesize,
    baselinestretch=1,
    breaklines}

\usepackage{tabularx}

\begin{document}

\def\sectionautorefname{Section}
\def\subsectionautorefname{Section}
\def\figureautorefname{Figure}

\newcommand{\bpstart}[1]{\smallskip\noindent{\textbf{#1.}}}
\newcommand{\code}[1]{{\fontfamily{phv}\selectfont\small {#1}}}

\newcommand{\colorcode}[1]{\mintinline[fontfamily=helvetica,bgcolor=blue!5,fontsize=\small]{python}{#1}}

\newcommand{\todo}[1]{{\color{orange}\bf [todo: {#1}]}}
\newcommand{\new}[1]{{ {#1}}}

\newsavebox{\fmbox}
\newenvironment{smpage}[1]
{\begin{lrbox}{\fmbox}\begin{minipage}{#1}}
{\end{minipage}\end{lrbox}\usebox{\fmbox}}

\newcommand*\circled[1]{\tikz[baseline=(char.base)]{
            \node[shape=circle,draw,inner sep=0.3pt] (char) {#1};}}

\firstsection{Introduction}

\maketitle

\begin{figure}[t]
    \centering
    \includegraphics[width=0.95\linewidth]{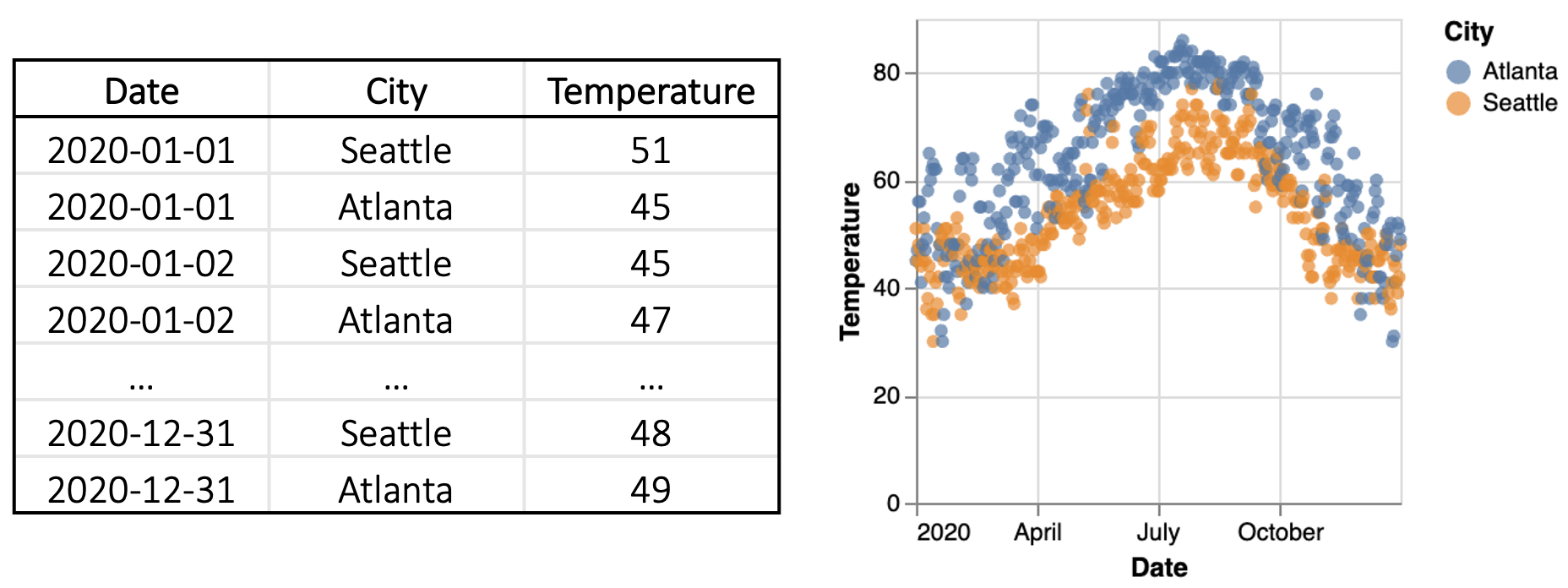}
    \caption{A dataset of Seattle and Atlanta daily temperatures in 2020 (left) and a scatter plot that visualizes them by mapping Date to $x$-axis, Temperature to $y$-axis, and City to color (right).}
    \label{fig:sea-atl-temp-simple}
\end{figure}

Most modern visualization authoring tools (e.g., Charticulator~\cite{ren2019charticulator}, Data Illustrator~\cite{liu2018data}, Lyra~\cite{satyanarayan2014lyra}) and libraries (e.g., ggplot2~\cite{wickham2009ggplot2}, Vega-Lite~\cite{satyanarayan2017vegalite}) expect tidy data~\cite{wickham2014tidy-data}, where every variable to be visualized is a column and each observation is a row. When the input data is in the tidy format, authors simply need to bind data columns to visual channels (e.g., \code{Date} $\mapsto x$-axis, \code{Temperature} $\mapsto y$-axis, \code{City} $\mapsto$ color in \cref{fig:sea-atl-temp-simple}). Otherwise, they need to prepare the data, even if the original data is clean and contains all information needed~\cite{bartram2021untidy}. Authors usually rely on data transformation libraries (e.g., tidyverse~\cite{wickham2019tidyverse}, pandas~\cite{the_pandas_development_team_2023_7741580}) or separate interactive tools (e.g., Wrangler~\cite{kandel2011wrangler}) to transform data into the appropriate format. However, authors need either programming experience or tool expertise to transform data, and they have to withstand the overhead of switching between visualization and data transformation steps. The challenge of data transformation remains a barrier in visualization authoring.

\begin{figure*}[ht]
    \centering
    \includegraphics[width=\linewidth]{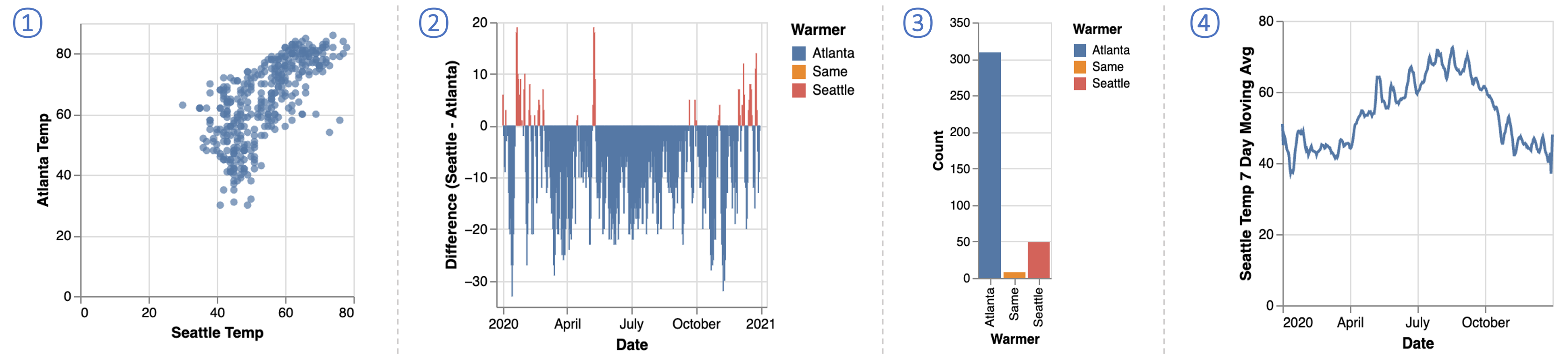}
    \caption{Visualizations created from \code{df} in \cref{fig:sea-atl-temp-simple} that require data transformation: (1) a scatter plot with Seattle and Atlanta temperatures on $x,y$-axes, (2) a bar chart to visualize the temperature difference between the two cities, (3) a histogram to count the number of days each city being warmer, and (4) a smoothed line chart that shows the 7-day moving averages of Seattle temperature.}
    \label{fig:sea-atl-temp-pivot-derived}
\end{figure*}

To address the data transformation challenge, we explore a fundamentally different approach for visualization authoring, leveraging an AI agent. We separate the high-level visualization intent ``\emph{what to visualize}'' from the low-level data transformation steps of ``\emph{how to format data to visualize},'' and automate the latter to reduce the data transformation burden. Specifically, we support two key types of data transformations (and their combinations) needed for visualization authoring:

\begin{itemize}
    \item \textbf{Reshaping}: A variable to be visualized is spread across multiple columns or one column includes multiple variables. For example, if authors want to create a different scatter plot from the table in \cref{fig:sea-atl-temp-simple} by mapping \code{Seattle} and \code{Atlanta} temperatures to $x,y$-axes (\cref{fig:sea-atl-temp-pivot-derived}-\circled{1}), they need to first ``pivot'' the table from long to wide format, because both variables of interest are stored in the \code{Temperature} column and are not readily available.
    \item \textbf{Derivation}: A variable needs to be extracted or derived from one or more existing columns. For example, if authors want to create a bar chart to show daily temperature differences between two cities (\cref{fig:sea-atl-temp-pivot-derived}-\circled{2}) and a histogram to count the number of days which city is warmer (\cref{fig:sea-atl-temp-pivot-derived}-\circled{3}), they need to derive the temperature difference and the name of the warmer city from the two cities' temperature columns, and map them to the $y$-axis and $x$-axis, respectively, and the city name to color channels of the corresponding charts. The derivation is also needed when the variable to be visualized requires analytical computation (e.g., aggregation, moving average, percentile) across multiple rows from a column in the table. For example, to plot a line chart to visualize the 7-day moving averages of \code{Seattle} temperatures (\cref{fig:sea-atl-temp-pivot-derived}-\circled{4}), the authors need to calculate the moving average using a window function and map it to  $y$-axis with \code{Date} on $x$-axis.
\end{itemize}

In this paper, we introduce Data Formulator, an interactive visualization authoring tool that embodies a new paradigm, \emph{concept binding}. To create a visualization with Data Formulator, authors provide their visualization intent by binding data concepts to visual channels. Upon loading of a data table, existing data columns are provided as known data concepts. When the required data concepts are not available to author a given chart, the authors can create the concepts: either using natural language prompts (for derivation) or by providing examples (for reshaping). Data Formulator handles these two cases differently, with different styles of input and feedback, and we provide a detailed description of how they are handled in \cref{sec:user-experience}. Once the necessary data concepts are available, the authors can select a chart type (e.g., scatter plot, histogram) and map data concepts to desired visual channels. If needed, Data Formulator dispatches the backend AI agent to infer necessary data transformations to instantiate these new concepts based on the input data and creates candidate visualizations. Because the authors' high-level specifications can be ambiguous and Data Formulator may generate multiple candidates, Data Formulator provides feedback to explain and compare the results. With this feedback, the authors can inspect, disambiguate, and refine the suggested visualizations. After that, they can reuse or create additional data concepts to continue their visualization authoring process.

We also report a chart reproduction study conducted with 10 participants to gather feedback on the new concept binding approach that employs an AI agent, and to evaluate the usability of Data Formulator. After an hour-long tutorial and practice session, most participants could create desired charts by creating data concepts—both with derivation and reshaping transformations. We conclude with a discussion on the lessons learned from the design and evaluation of Data Formulator, as well as important future research directions.
\section{\new{Illustrative Scenarios}}\label{sec:user-experience}
In this section, we illustrate users' experiences to create visualizations in \cref{fig:sea-atl-temp-simple,fig:sea-atl-temp-pivot-derived} using programs and Data Formulator from the initial input data in \cref{fig:sea-atl-temp-simple}. We refer to this dataset as \code{df} in this section.

\subsection{Experience with Programming}

We first illustrate how an experienced data scientist, Eunice, uses programming to create the desired visualizations with pandas and Altair libraries in Python.

\bpstart{Daily Temperature Trends} Eunice starts with the scatter plot in \cref{fig:sea-atl-temp-simple}. Because \code{df} is in the tidy format with \code{Date}, \code{City}, and \code{Temperature} available, Eunice needs no data transformation and writes a simple Altair program to create the plot:

\begin{center}
\begin{smpage}{0.9\linewidth}
\begin{minted}[fontfamily=helvetica,fontsize=\small]{python}
alt.Chart(df).mark_circle().encode(x='Date', y='Temperature', color='City')
\end{minted}
\end{smpage}
\end{center}
This program calls the Altair library (\code{alt}), selects the input dataset \code{df} and the scatter plot function \code{mark\_circle}, and maps columns to $x,y$ and color channels. It renders the desired scatter plot in \cref{fig:sea-atl-temp-simple}.
 
\begin{figure}[t]
    \centering
    \includegraphics[width=0.9\linewidth]{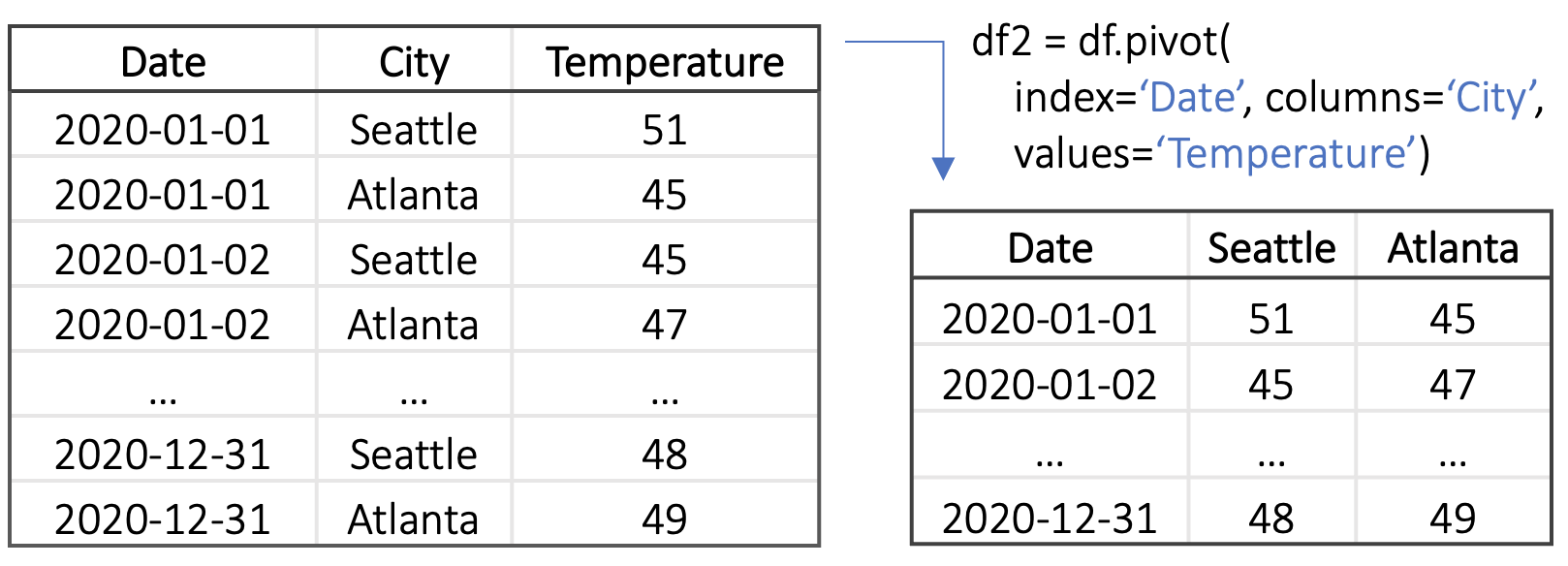}
    \caption{Prepare the new data \code{df2} with the \code{pivot} function to populate \code{Seattle} and \code{Atlanta} temperatures from \code{City} and \code{Temperature} columns.}
    \label{fig:pivot-table}
\end{figure}

\bpstart{Seattle vs. Atlanta Temperatures} To make a more direct comparison of two cities' temperatures, Eunice wants to create a different scatter plot (\cref{fig:sea-atl-temp-pivot-derived}-\circled{1}) by mapping \code{Seattle} and \code{Atlanta} temperatures to $x,y$-axes. However, \code{Seattle} and \code{Atlanta} temperatures are not available as columns in \code{df}. She therefore needs to transform \code{df} to surface them. Because \code{df} is in the ``long'' format, where temperatures of both cities are stored in one column \code{Temperature}, she needs to pivot the table to the ``wide'' format. Eunice switches to the data transformation step and uses the \code{pivot} function from the pandas library to reshape \code{df} (\cref{fig:pivot-table}). This program populates \code{Seattle} and \code{Atlanta} as new column names from the \code{City} column, and their corresponding \code{Temperature} values are moved to these new columns by \code{Date}. 
With \code{df2}, Eunice creates the desired visualization, which maps \code{Seattle} and \code{Atlanta} to $x,y$-axes of the scatter plot with the following program:

\begin{center}
\begin{smpage}{0.9\linewidth}
\begin{minted}[fontfamily=helvetica,fontsize=\small]{python}
alt.Chart(df2).mark_circle().encode(x='Seattle', y='Atlanta')
\end{minted}
\end{smpage}
\end{center}

\begin{figure}[t]
    \centering
    \includegraphics[width=0.9\linewidth]{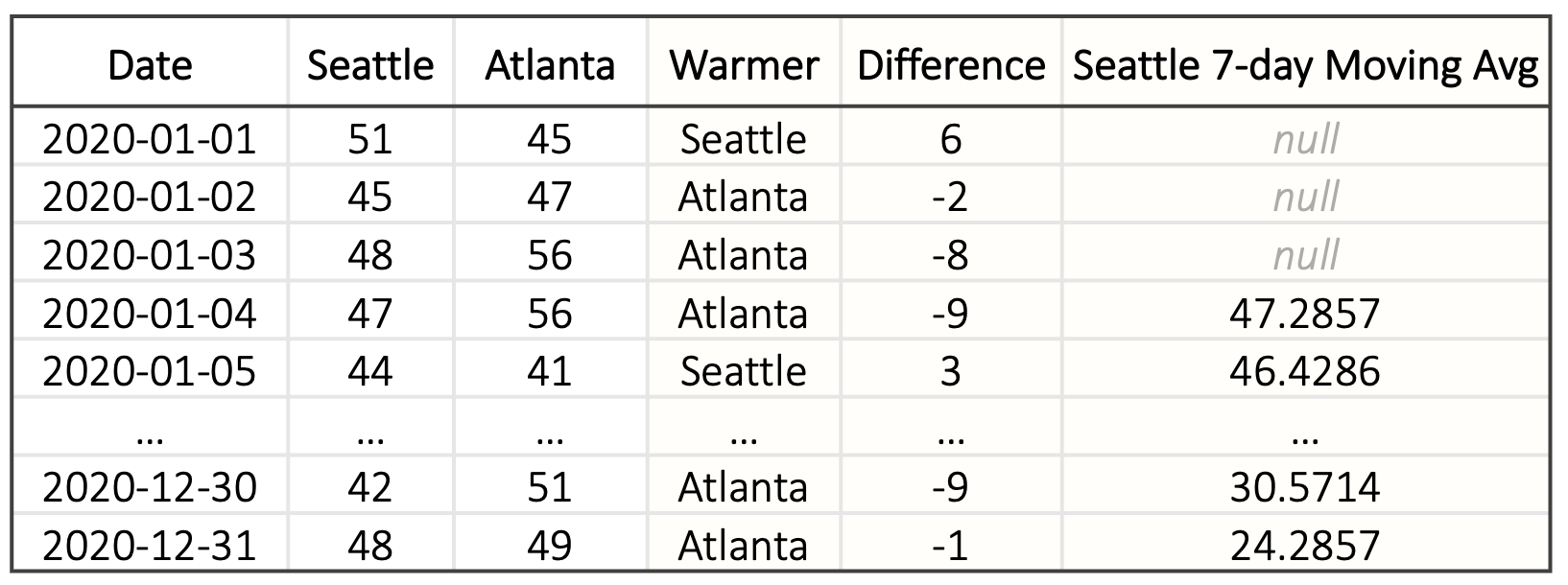}
    \caption{Extend \code{df2} in \cref{fig:pivot-table} to derive \code{Warmer}, \code{Difference}, and \code{Seattle 7-day Moving Avg} columns that are necessary for visualizations in \cref{fig:sea-atl-temp-pivot-derived}.}
    \label{fig:derived-table}
\end{figure}

\bpstart{Temperature Differences} Eunice wants to create two visualizations to show how much warmer is Atlanta compared to Seattle: a bar chart to visualize daily temperate differences (\cref{fig:sea-atl-temp-pivot-derived}-\circled{2}) and a histogram to show the number of days each city is warmer (\cref{fig:sea-atl-temp-pivot-derived}-\circled{3}). Again, because necessary fields \code{Difference} and \code{Warmer} are not in \code{df2}, Eunice needs to transform the data. This time, she writes a program to perform column-wise computation, which extends \code{df2} with two new columns \code{Warmer} and \code{Difference} (\cref{fig:derived-table}). 
Eunice then creates the daily temperature differences chart by mapping \code{Date} and \code{Difference} to $x,y$-axes and the histogram by mapping \code{Warmer} to $x$-axis and the aggregation function, \code{count()}, to $y$-axis to calculate the number of entries.

\begin{center}
\begin{smpage}{0.85\linewidth}
\begin{minted}[fontfamily=helvetica,fontsize=\small]{python}
# extend df2 with new columns 'Difference' and 'Warmer'
df2['Difference'] = df2['Seattle'] - df2['Atlanta']
df2['Warmer'] = df2['Difference'].apply(
    lambda x: 'Seattle' if x > 0 else ('Atlanta' if x < 0 else 'Same'))

# create the bar chart
alt.Chart(df2).mark_bar().encode(x='Date', y='Difference', color='Warmer')
# create the histogram
alt.Chart(df2).mark_bar().encode(x='Warmer', y='count()', color='Warmer')
\end{minted}
\end{smpage}
\end{center}

\begin{figure*}[t]
    \includegraphics[width=\linewidth]{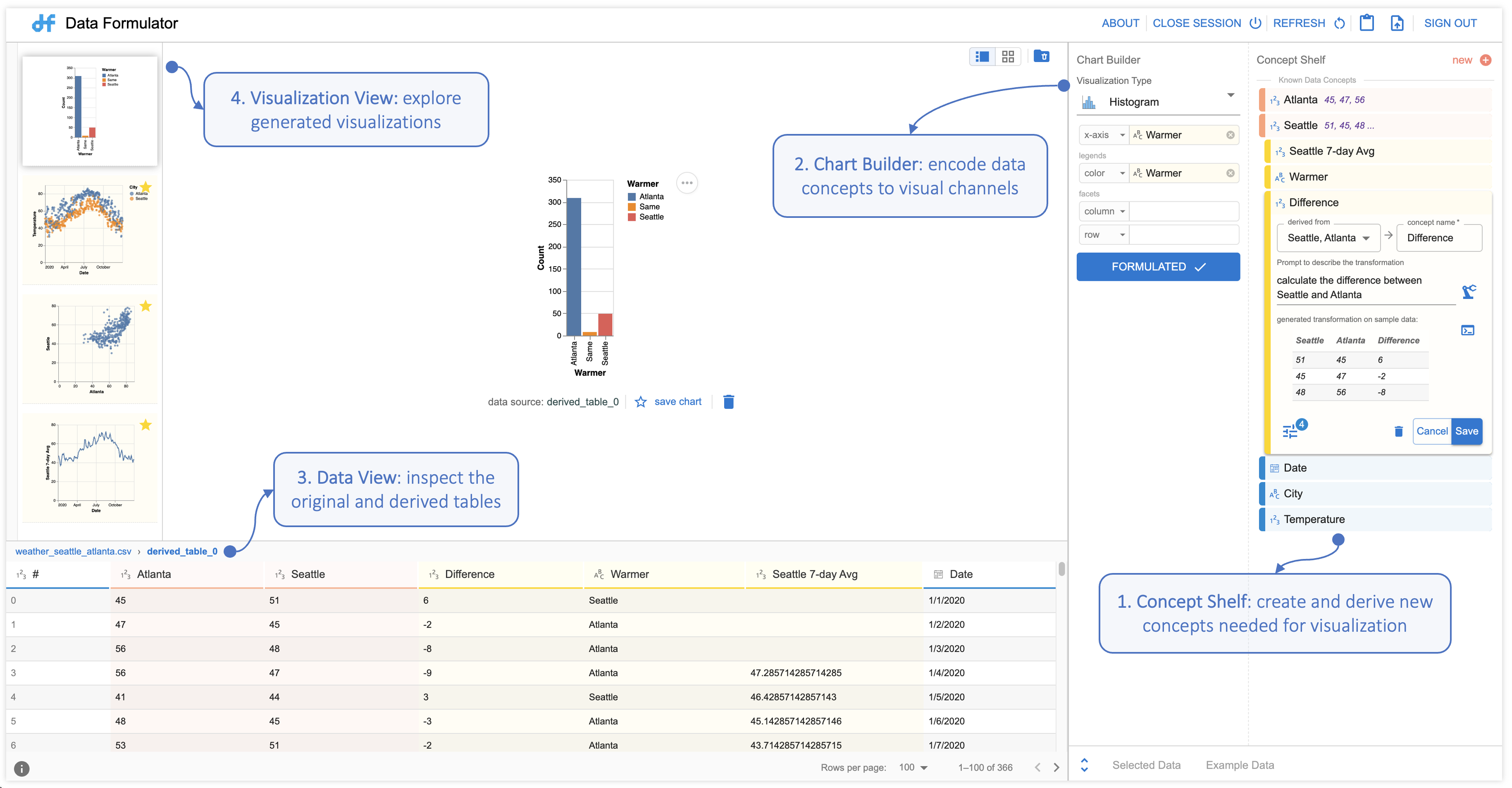}
    \caption{Data Formulator UI. After loading the input data, the authors interact with Data Formulator in four steps: (1) in the Concept Shelf, create (e.g., \code{Seattle} and \code{Atlanta}) or derive (e.g., \code{Difference}, \code{Warmer}) new data concepts they plan to visualize, (2) encode data concepts to visual channels of a chart using Chart Builder and formulate the chart, (3) inspect the derived data automatically generated by Data Formulator, and (4) examine and save generated visualizations. Throughout the process, Data Formulator provides feedback to help authors understand generated data and visualizations.}
    \label{fig:data-formulate-ui}
\end{figure*}

\bpstart{7-day Moving Average of Seattle's Temperature} Finally, Eunice wants to include a line chart for Seattle temperature trends in the report. Because daily temperatures fluctuate, she decides to create a smooth line chart based on 7-day moving average temperatures. Eunice needs an analytical function to calculate the moving average. Because the input data is sorted by \code{Date}, Eunice chooses the \code{rolling} function from pandas: she sets \code{window=7} and \code{center=True} so that the moving average is calculated with a sliding window from day $d-3$ to day $d+3$ for each date $d$.
This transformation adds the new column \code{Seattle 7-day Moving Avg} to \code{df2} (\cref{fig:derived-table}; the first 3 days are null because of insufficient data), and Eunice maps \code{Date} and the new column to a line chart to create the desired visualization (\cref{fig:sea-atl-temp-pivot-derived}-\circled{4}).

\begin{center}
\begin{smpage}{0.95\linewidth}
\begin{minted}[fontfamily=helvetica,fontsize=\small]{python}
df2['Seattle 7-day Moving Avg'] = df2['Seattle'].rolling(window=7, center=True)
alt.Chart(df2).mark_line().encode(x='Date', y='Seattle 7-day Moving Avg')
\end{minted}
\end{smpage}
\end{center}

\bpstart{Remark} In all cases, Eunice can specify visualizations using simple Altair programs by mapping data columns to visual channels. However, data transformation steps make the visualization process challenging. Eunice needs to choose the right type of transformation based on the input data and desired visualization (e.g., creating the scatter plot in \cref{fig:sea-atl-temp-simple} from \code{df2} would require unpivot instead). Furthermore, Eunice needs knowledge about \code{pandas} to choose the right function and parameters per task (e.g., \code{rolling} will not fit if Eunice wants to calculate moving average for each city in \code{df}). Eunice's programming experience and data analysis expertise allowed her to successfully complete all tasks. But a less experienced data scientist, Megan, finds this process challenging. Megan decides to use Data Formulator to reduce the data transformation overhead.

\subsection{Experience with Data Formulator} 

Data Formulator (\cref{fig:data-formulate-ui}) has a similar interface as ``shelf-configuration''-style visualization tools like Tableau or Power BI. But unlike these tools that support only mappings from input data columns to visual channels, Data Formulator enables authors to create and derive new data concepts and map them to visual channels to create visualizations \emph{without requiring manual data transformation}.

\bpstart{Daily Temperature Trends} Once Megan loads the input data (\cref{fig:sea-atl-temp-simple}), Data Formulator populates existing data columns (\code{Date}, \code{City}, and \code{Temperature}) as known \emph{data concepts} in the Concept Shelf. Because all three data concepts are already available, no data transformation is needed. Megan selects the visualization type ``Scatter Plot'' and maps these data concepts to $x,y$ and color channels in Chart Builder through drag-and-drop interaction. Data Formulator then generates the desired scatter plot. 

\begin{figure*}[t]
    \centering
    \includegraphics[width=\linewidth]{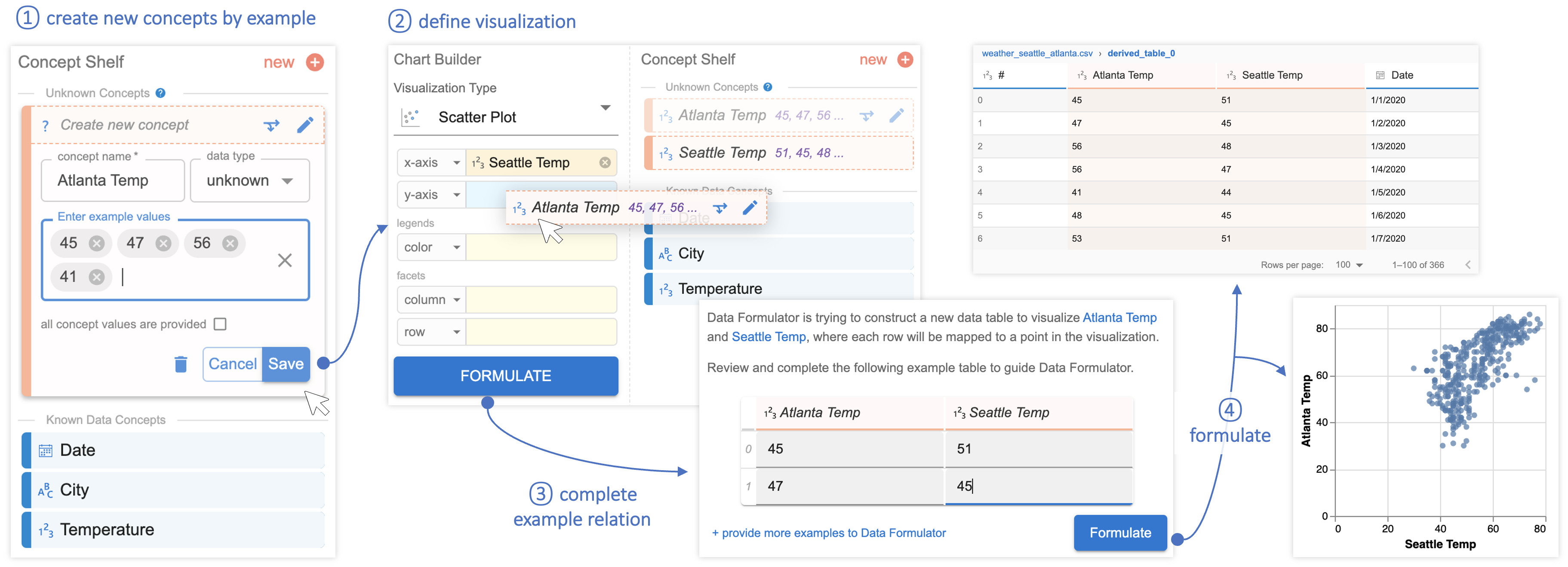}
    \caption{Megan (1) creates new data concepts, \code{Seattle Temp} and \code{Atlanta Temp}, by providing examples and (2) maps them to $x,y$-axes of a scatter plot to specify the visualization intent. (3) Data Formulator asks Megan to provide a small example to illustrate how these two concepts are related, and Megan confirms the example. (4) Based on the example, Data Formulator generates the data transformation and creates the desired visualization.}
    \label{fig:data-formulator-pivot}
\end{figure*}

\begin{figure*}[t]
    \centering
    \includegraphics[width=\linewidth]{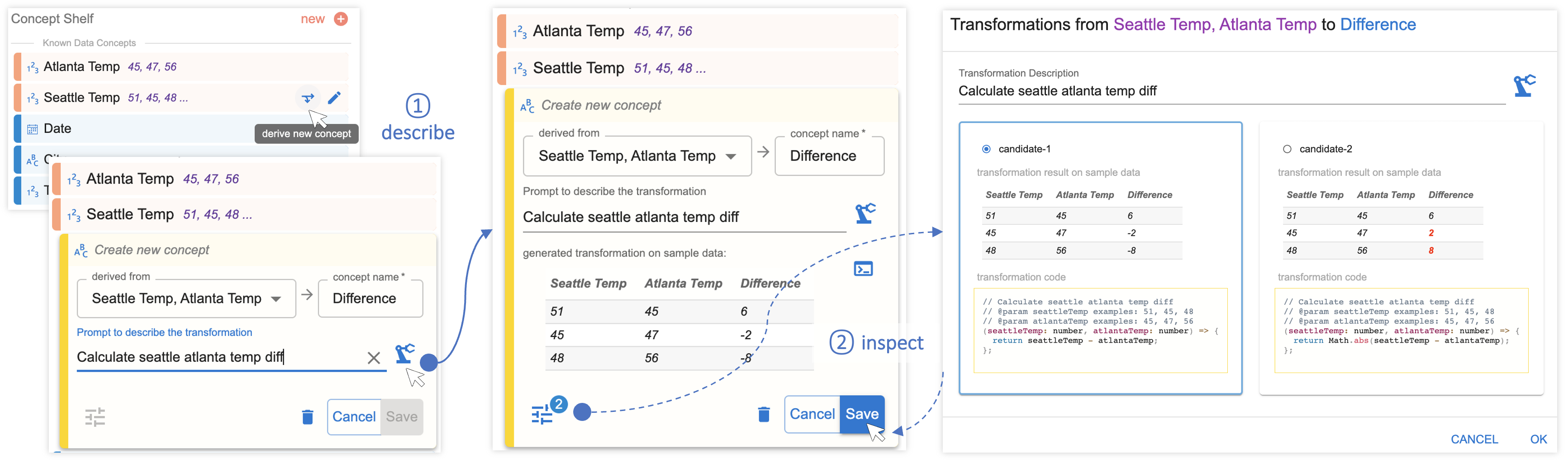}
    \caption{(1) Megan derives the new concept \code{Difference} from \code{Atlanta Temp} and \code{Seattle Temp} using natural language. Data Formulator generates two candidates and displays the first one in the concept card. (2) Megan opens the dialog to inspect both, confirms the first one, and saves the concept.}
    \label{fig:data-formulator-derive}
\end{figure*}

\begin{figure}[t]
    \centering
    \includegraphics[width=\linewidth]{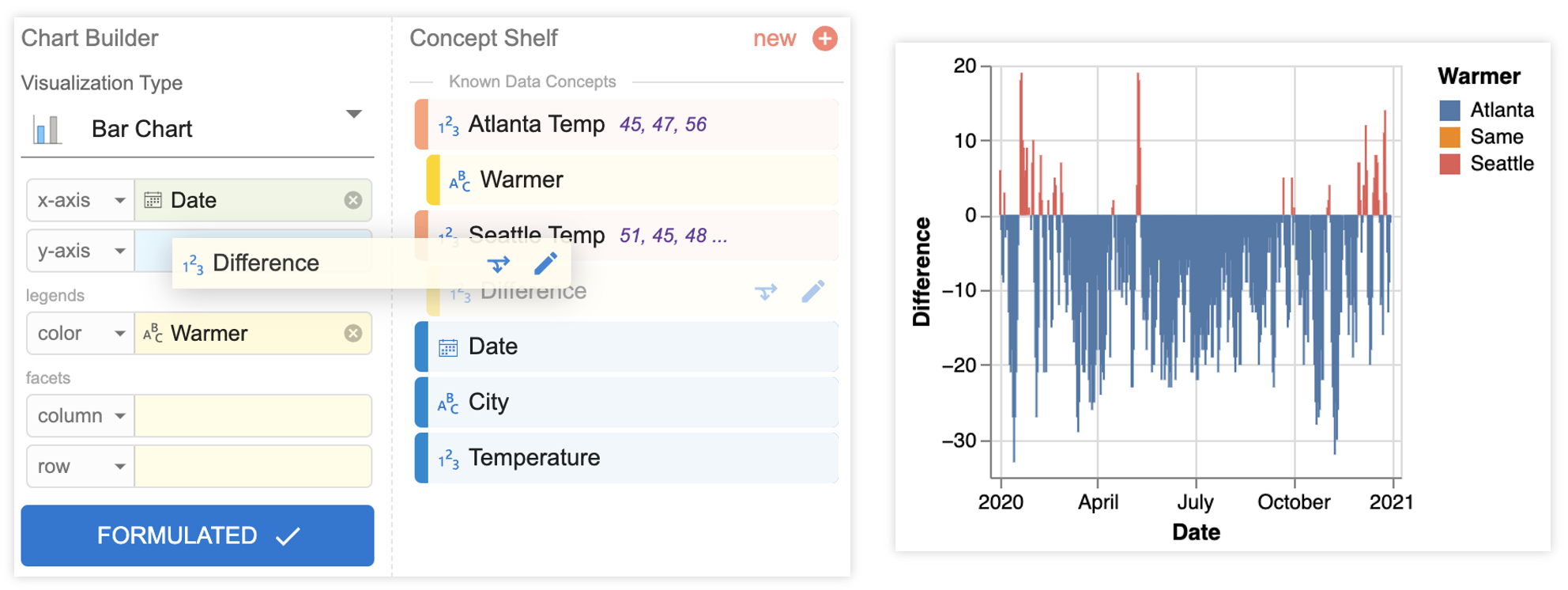}
    \caption{Megan creates the bar chart using derived concepts, Difference and Warmer, as well as an original concept Date.}
    \label{fig:data-formulator-derive-bar-chart}
\end{figure}

\begin{figure}[t]
    \centering
    \includegraphics[width=\linewidth]{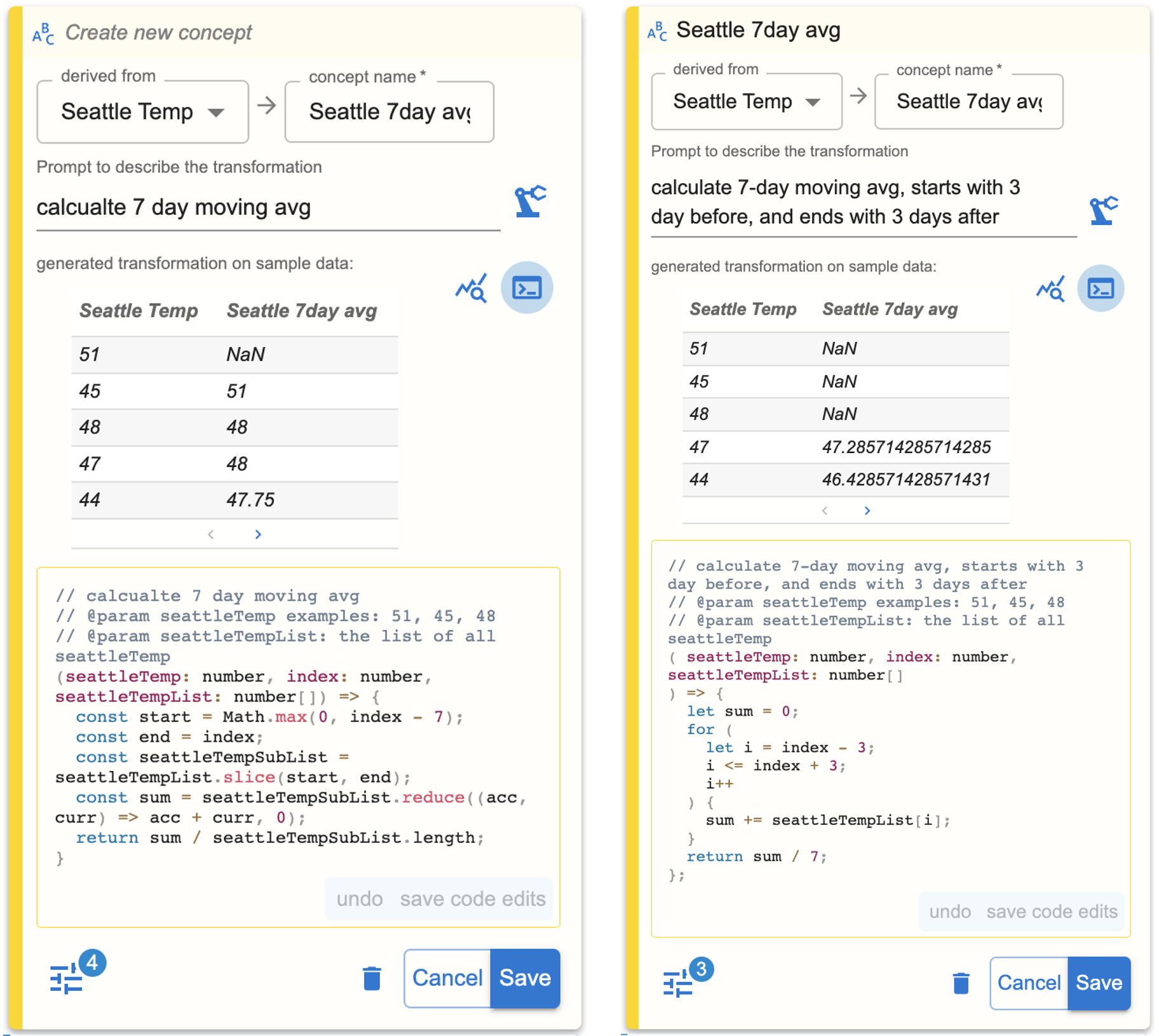}
    \caption{Megan derives the 7-day moving averages from Seattle Temp. After inspecting the results, she edits the description to be more precise.}
    \label{fig:data-formulator-derive-7-day-avg}
\end{figure}

\bpstart{Seattle vs. Atlanta Temperatures} To create the second scatter plot (\cref{fig:derived-table}-\circled{1}), Megan needs to map \code{Seattle} and \code{Atlanta} temperatures to $x,y$-axes of a scatter plot. Because \code{Seattle} and \code{Atlanta} temperatures are not available as concepts yet, 
Megan starts out by creating a new data concept \code{Atlanta Temp} (\cref{fig:data-formulator-pivot}-\circled{1}): she clicks the {\sf\small\color{orange} new \circled{+}} button in the Concept Shelf, which opens a concept card that asks her to name the new concept and provide some examples values; Megan provides four Atlanta temperatures (45, 47, 56, 41) from the input data as examples and saves it. Similarly, Megan creates another new concept \code{Seattle Temp}. Because Data Formulator's current knowledge to them is limited to their names and example values, both concepts are listed as an unknown concept for now. (They will be resolved later when more information is provided.) 

With these new concepts and the Scatter Plot selected, 
Megan maps new data concepts \code{Seattle Temp} and \code{Atlanta Temp} to $x,y$-axes (\cref{fig:data-formulator-pivot}-\circled{2}), and then clicks the FORMULATE button to let Data Formulator formulate the data and instantiate the chart. Based on the visualization spec, Data Formulator realizes that the two unknown concepts are related to each other but not yet certain how they relate to the input data. Thus, Data Formulator prompts Megan with an example table to complete: each row in the example table will be a data point in the desired scatter plot. Megan needs to provide at least two data points from the input data to guide Data Formulator on how to generate this transform (\cref{fig:data-formulator-pivot}-\circled{3}). Here, Megan provides the temperatures of Atlanta and Seattle on 01/01/2020 and 01/02/2020 from the table \cref{fig:sea-atl-temp-simple}. When Megan submits the example, Data Formulator infers a program that can transform the input data to generate a new table with fields \code{Atlanta Temp} and \code{Seattle Temp} that subsumes the example table provided by Megan. Data Formulator generates the new table and renders the desired scatter plot (\cref{fig:data-formulator-pivot}-\circled{4}). Megan inspects the derived table and visualization and accepts them as correct.

\bpstart{Temperature Differences} To create a bar chart and a histogram to visualize temperature differences between the two cities, Megan needs two new concepts, \code{Difference} and \code{Warmer}. 
This time, Megan notices that both concepts can be \emph{derived} from existing fields based on column-wise mappings, and thus she uses the ``derive'' function of Data Formulator (\cref{fig:data-formulator-derive}). Megan first clicks the ``derive new concept'' option on the existing concept \code{Seattle Temp}, which opens up a concept card that lets her describe the transformation she wants using natural language. Megan selects \code{Seattle Temp} and \code{Atlanta Temp} as the ``derived from'' concepts, provides a name \code{Difference} for the new concept, and describes the transform using natural language, ``Calculate seattle atlanta temp diff.'' Megan then clicks the generate button and Data Formulator dispatches its backend AI agent to generate code. Data Formulator returns two code candidates and presents the first one in the concept card. Megan opens up the dialog to inspect both candidates and learns that because her description did not clearly specify whether she wants the difference or its absolute value, Data Formulator returns both options as candidates. After inspecting the example table and the transformation code provided by Data Formulator, Megan confirms the first candidate and saves the concept \code{Difference}. Similarly, Megan creates a concept, \code{Warmer}, from \code{Seattle Temp} and \code{Atlanta Temp} with the description ``check which city is warmer, Atlanta, Seattle, or same.'' Data Formulator applies the data transformation on top of the derived table from the last task and displays the extended table in Data View (\cref{fig:data-formulate-ui}). Because both concepts are now ready to use, Megan maps them to Chart Builder to create the desired visualizations (\cref{fig:data-formulator-derive-bar-chart}).

\bpstart{7-day Moving Average of Seattle's Temperature} Last, Megan needs to create a line chart with 7-day moving average temperatures. Because the moving average can be derived from the \code{Seattle Temp} column, Megan again chooses to use the derive function.
Megan starts with a brief description ``calculate 7-day moving avg'' and calls Data Formulator to generate the desired transformation. Upon inspection, Megan notices that the generated transformation is close but does not quite match her intent: the 7-day moving average starts from $d-6$ to $d$ for each day $d$ as opposed to $d-3$ to $d+3$ (\cref{fig:data-formulator-derive-7-day-avg}). Based on this observation, Megan changes the description into ``calculate 7-day moving avg, starts with 3 days before, and ends with 3 days after'' and re-runs Data Formulator. This time, Data Formulator generates the correct transformation and presents the extended data table in \cref{fig:data-formulate-ui}. Megan then maps \code{Date} and {\footnotesize\sf Seattle 7-day Moving Avg} to $x,y$-axes of a line chart.

\bpstart{Remark} With the help of Data Formulator, Megan creates visualizations without manually transforming data. Instead, Megan specifies the data concepts she wants to visualize by:
\begin{itemize}\itemsep-5pt
\item building new concepts using examples (when the new concept is spread among multiple columns or multiple concepts are stored in the same column, e.g., \code{Seattle Temp} and \code{Atlanta Temp} are both stored in the  \code{Temperature} column); and
\item deriving new concepts using natural language (when the new concept can be computed from existing ones using column-wise operators, e.g., \code{Difference} from \code{Seattle Temp} and \code{Atlanta Temp}).
\end{itemize}

Megan then drags-and-drops data concepts to visual channels of a chart. In this process, for derived concepts, Data Formulator displays generated candidate code and example table to help Megan inspect and select the transformation; for concepts created by example, Data Formulator prompts Megan to elaborate their relations by completing an example table. Data Formulator then transforms the data and generates the desired visualizations. Data Formulator reduces Megan's visualization overhead by shifting the task of specifying data transformation into the task of inspecting generated data. Because Data Formulator's interaction model centers around data concepts, Megan does not need to directly work with table-level operators, such as \code{pivot}, {\footnotesize\sf map/reduce} and \code{partitioning}, which are challenging to master.

\section{The Data Formulator Design}
\label{sec:design}

In this section, we describe our design principles, explain Data Formulator's interaction model, and how Data Formulator derives data concepts and formulates visualizations from the author's inputs.

\subsection{Design Principles}
Data Formulator introduces \emph{data concepts}, an abstraction of the columns needed for an author to specify their target visualization. To eliminate the author's burden to manually transform the data table before plotting, we designed Data Formulator based on the following guiding design principles.

\bpstart{Treat design concepts as first-class objects} The notion of data concepts is a generalization of table columns: it is a reference to columns both from a current table and from a future transformed table. They offer two benefits. First, concept-level transformations are easier to describe and understand than table-level operators. Table-level transformations require either advanced operators like \code{pivot} and \code{unpivot}, or high-order functions like {\sf\footnotesize map} and \code{window}, while concept-level operators are first-order functions over primitive elements (e.g., arithmetic) or lists (e.g., percentile). This makes it easier for the author to communicate with the AI agent and verify the results. Second, we can build the interaction experience on top of existing designs people are already familiar with: data concepts resemble data columns existing shelf-configuration tools commonly use.

\bpstart{Leverage benefits from multiple interaction approaches} Data Formulator employs both natural language interaction (for deriving concepts) and programming-by-example approach (for building custom concepts). Natural language descriptions have a superior ability to translate high-level intent into executable code and large language models (LLMs) can reason about natural concepts (e.g., academic grades are A, B, C, D, and F; months are from January to December). However, it can be difficult for the author to provide proper descriptions if they do not understand notions like pivoting, and natural language descriptions can be imprecise and ambiguous. In contrast, while program synthesizers cannot reason about natural concepts, they are less ambiguous, and it is easier for the author to convey reshaping operations by demonstrating the output relation. By incorporating multiple approaches and feedback for different transformation types (derivation vs. reshaping), Data Formulator takes advantage of both, reducing the specification barrier and improving the likelihood for the AI agent to generate correct and interpretable codes.

\bpstart{Ensure correct data transformation and promote trust} While LLM and program synthesizers can automatically generate code to eliminate the author's manual data transformation burden, they can incorrectly generalize the author's specification. Therefore, it is crucial for the author to view and verify the results. Our design employs mechanisms to ensure such inspection by the author: (1) display multiple candidates for the author to review, if available, (2) display both the code (process) and the sample output values (results) to help the author understand the transformation, and (3) allow the author to edit the generated transformation code to correct or refine it.

\bpstart{Improve the system expressiveness} Data Formulator's expressiveness is defined by the combination of transformation function and visualization language. Data Formulator's visualization spec builds on top of Vega-Lite specifications. While Data Formulator's UI does not provide options to layer marks, the author can import their custom Vega-Lite specs of layered visualizations to achieve the same design. For data transformation, Data Formulator supports reshaping options from tidyverse as described in \cref{sec:user-experience}, and it supports both column-wise derivation and analytical computation that can be generated by the LLM. Note that while our transformation language does not include aggregation, the author can achieve the same visualization by setting aggregation options on the desired axes (e.g., map \code{Month} to $x$-axis and \code{avg(Seattle Temp)} to $y$-axis to create a bar chart with average temperature). However, with the current design, the author cannot derive or reshape data that first require aggregation without re-importing the aggregated data.

\subsection{Interaction Model}

\begin{figure}[h]
    \centering
    \includegraphics[width=\linewidth]{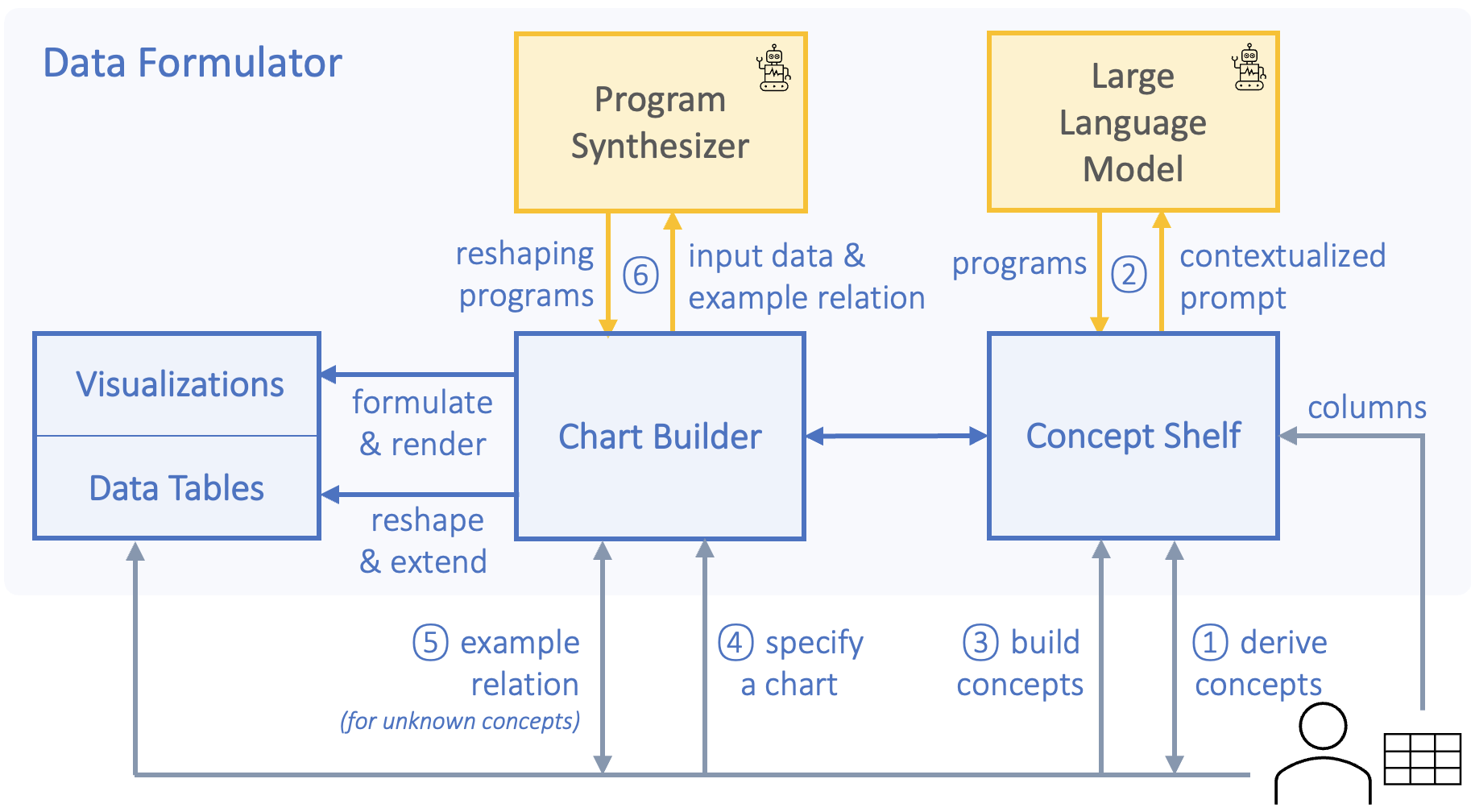}
    \caption{Data Formulator's interaction model.}
    \label{fig:design-program-overview}
\end{figure}

\Cref{fig:design-program-overview} shows Data Formulator's high-level interaction model.
Data Formulator first loads data columns from the input table as original (and known) concepts (e.g., \code{Date}, \code{City}, and \code{Temperature} concepts in \cref{fig:data-formulate-ui}). 
The author uses the Concept Shelf to create new data concepts, if needed, in two ways (\cref{sec:create-concepts}): (1) derive a concept from existing ones by interacting with an AI agent using natural language or (2) build a custom concept by providing example values. If the new concept is derived from known concepts, Data Formulator immediately extends the current data table and registers it as a known concept.

With necessary data concepts known, the author uses the Chart Builder to map data concepts to visual channels of a chart. If unknown custom concepts are used to specify a visualization, Data Formulator asks the author to provide an example relation among the encoded concepts to transform the input table by using a programming-by-example approach.
With the necessary data formulations applied, Data Formulator generates a Vega-Lite spec and renders the visualization. 

\subsection{Creating New Data Concepts}
\label{sec:create-concepts}

The author can \textbf{derive} a concept from one or more data concepts by interacting with Data Formulator's AI agent (\cref{fig:design-program-overview}-\circled{1}).
In addition to a concept name, the author provides both a list of source concepts from which the new concept is derived and a natural language description of the transformation (\cref{fig:data-formulator-derive}-\circled{1}). Data Formulator then generates a contextualized prompt that grounds the description in the context of source concepts. 
This prompt combines the author's description and the descriptions of input parameters for all source concepts (with example values sampled from their domains) as comments, and joins it with the function prefix to instruct the AI agent to complete a Typescript function (as opposed to generate non-code text or uncontrolled code snippets). Data Formulator prepares two types of prompts for each query to cover simple derivation (Example 1) and analytical computation (Example 2) because it does not know if analytical computation is needed beforehand.

\vspace{1mm}
\noindent \textbf{Example 1:} The prompt for ``Calculate seattle atlanta temp diff'' with source concepts \code{Seattle Temp} and \code{Atlanta Temp} (\cref{fig:data-formulator-derive}).
\begin{center}
\begin{smpage}{0.7\linewidth}
\begin{minted}[fontfamily=helvetica,fontsize=\small]{typescript}
// Calculate seattle atlanta temp diff
// @param seattleTemp examples: 51, 45, 48
// @param atlantaTemp examples: 45, 47, 56
(seattleTemp: number, atlantaTemp: number) => {
\end{minted}
\end{smpage}
\end{center}

\noindent \textbf{Example 2:} The prompt for ``calculate 7-day moving avg'' with source concept \code{Seattle Temp} (\cref{fig:data-formulator-derive-7-day-avg}). It provides \code{index} and \code{seattleTempList} so that the function can access to other values of the \code{seattleTemp} when analytical computation is needed (e.g., calculate the moving average for current index, derive percentile of the seatteTemp among all values).
\begin{center}
\begin{smpage}{0.95\linewidth}
\begin{minted}[fontfamily=helvetica,fontsize=\small]{typescript}
// calculate 7-day moving avg
// @param seattleTemp examples: 51, 45, 48
// @param seattleTempList: the list of all seattleTemp
(seattleTemp: number, index: number, seattleTempList: number[]) => { 
\end{minted}
\end{smpage}
\end{center}

Data Formulator sends both prompts to LLM (we use Codex Davinci 2~\cite{chen2021evaluating}) to generate the transformation code (\cref{fig:design-program-overview}-\circled{2}), asking for five candidate completions. When candidate programs are returned from LLM, Data Formulator filters out programs that are not executable or contain error outputs by executing them on sample values from source domains. Data Formulator then presents the programs along with their example execution results for the author to inspect (\cref{fig:data-formulator-derive}). Once confirmed, a new derived concept is created and shown in the Concept Shelf. If all source fields are known concepts, Data Formulator derives a new column by applying the transformation function to every tuple from the source columns and appends the column in the current table for the author to review (e.g., \cref{fig:data-formulate-ui}).


The author can also \textbf{build} a custom concept by providing its name and a set of example values that belong to its domain (\cref{fig:design-program-overview}-\circled{3}). Custom concepts are designed to support data reshaping: the author creates custom concepts when (1) the concept is spread across multiple columns; the author wants to combine multiple columns in a wide table to create one new concept in a long table, (2) multiple concepts are stored in one column; they want to surface fields from a long table, and (3) multiple values for a concept are collapsed in a column as a list (e.g., the value for an ``actors'' column is a list of actors for each movie); the author wants to split the list into multiple rows (i.e., one actor per row). 
These custom concepts are \textit{not} known yet upon creation because Data Formulator needs additional information from the author to resolve their relation with the input data. As we will describe in the next section, the resolution is achieved by inferring the reshaping program based on the  example relations provided by the user.

With data concepts (including newly crated ones) ready, the author is ready to interact with the Chart Builder to create visualizations.

\subsection{Specifying and Formulating the Visualization} 

Chart Builder employs a shelf-configuration interface: authors drag-and-drop data concepts to visual channels of the selected visualization to specify visual encoding. Based on the encoding, Data Formulator generates a Vega-Lite specification (e.g., \cref{fig:design-vl-specs}) to render the visualization. Data Formulator adopts a chart-based specification: each chart type corresponds to a Vega-Lite template with placeholder encodings to be filled from the author specification. Data Formulator currently supports scatter plots (circle-based, bubble chart, ranged dot plots), bar charts (single-column, stacked, layered, grouped, histogram), line charts (with and without dots), heatmap, and custom charts (with all compatible visual channels). 

\begin{figure}[h]
\centering
\begin{smpage}{\linewidth}
\begin{minted}[fontfamily=helvetica,fontsize=\small]{json}
{ "mark": "circle", "encoding" : { "x": {"field": "Date", "type": "temporal"}, "y": {"field": "Temperature", "type": "quantitative"},  "color": {"field": "City"} } }

{ "mark": "circle",  "encoding" : { "x": {"field": "Seattle Temp", "type": "quantitative"}, "y": {"field": "Atlanta Temp", "type": "quantitative"} } }
\end{minted}
\end{smpage}
\caption{Vega-Lite specs for the scatter plots in \cref{fig:sea-atl-temp-pivot-derived}-1 and \cref{fig:data-formulator-pivot}.}
\label{fig:design-vl-specs}
\end{figure}

When all fields used in the visual encoding are available, Data Formulator combines the Vega-Lite spec with the input data to render the visualization (e.g., \cref{fig:sea-atl-temp-simple}). Otherwise, when some concepts are unknown (unresolved custom concepts or concepts derived from unknown ones), Data Formulator first interacts with the author and then calls the program synthesis engine to create the transformed table. 

Once the author specifies the visual encoding, Data Formulator first checks if any unknown concepts are used. If so, it asks the author to illustrate the relation of unknown concepts with other concepts used in the visual encoding by filling out an example relation in a sample table (e.g., \cref{fig:design-program-overview}-\circled{5}). Data Formulator needs such example relation to disambiguate the visualization intent because unknown concepts contain example values only from their own domains, missing information on how they will be related row-wise in the transformed table. For example, Data Formulator generates the example relation with \code{Seattle Temp} and \code{Atlanta Temp} fields as shown in \cref{fig:data-formulator-pivot}-\circled{3} for the author to complete. To reduce the author's efforts, Data Formulator pre-fills two example values of \code{Atlanta Temp} based on its sample domain and asks the author to complete their corresponding \code{Seattle Temp} values (e.g., what's \code{Seattle Temp} when \code{Atlanta Temp} is 45). Each row in the example relation will be a row in the transformed data, which will then be mapped to a point in the scatter plot.

Once the author submits the example relation, Data Formulator calls the program synthesizer to solve the data reshaping problem (\cref{fig:design-program-overview}-\circled{6}). Given an example relation $E$, with input data $T$, the program synthesizer solves the programming-by-example problem to find a reshaping program $p$ such that $E\subseteq p(T)$ (i.e., the transformed data should generalize the example $E$). The reshaping program $p$ is defined by the grammar in \autoref{fig:reshaping-operators}, where $p$ is recursively defined over four core reshaping operators from the R tidyverse library. We include only reshaping operators because other operators like \code{unite} and \code{summarise} are already supported by Data Formulator's ability to derive concepts from natural language. With this grammar, the program synthesizer performs an enumerative search in the program space for candidate programs. To speed up this combinatorial search process, we leverage abstract interpretation to prune the search space: the program synthesis engine returns candidate programs that satisfy the example relation to Chart Builder. Note that multiple candidates could be generated since the example relation is small and potentially ambiguous. In practice, unlike other programming-by-example tools, the small example relation is precise enough to constrain the program space that only the correct candidate is returned, because the program synthesizer only needs to solve the reshaping problem.

\begin{figure}
\[\arraycolsep=1.6pt
\begin{array}{rcll}
    p & \leftarrow &  T \\
      & | & \mathsf{pivot\_longer}(p, \bar{c}) & {\color{gray}\textit{(pivot from wide to long)}}\\
      & | & \mathsf{pivot\_wider}(p, c_{name}, c_{vals}) & {\color{gray}\textit{(pivot from long to wide)}}\\
      & | & \mathsf{separate}(p, c) &  {\color{gray}\textit{(split a column into two)}}\\
      & | & \mathsf{separate\_rows}(p, c) &  {\small\color{gray}\textit{(separate a collapsed column into rows)}}
\end{array}
\]
\vspace{-10pt}
\caption{Reshaping operators supported by Data Formulator. $T$ refers to input data, and $c$ refers to column names.}
\label{fig:reshaping-operators}
\end{figure}

With generated reshaping programs, Chart Builder prepares the input data: it first generates a reshaped table from each reshaping program and then for every derived concept used in the encoding, it extends the reshaped table with a new column by applying the transformation function on every tuple in the table. This way, Data Formulator generates a new table with all necessary fields to instantiate the visualization. 

Data Formulator presents the prepared table and candidate visualizations for the author to inspect (\cref{fig:data-formulate-ui}-\circled{3}\circled{4}). When the author confirms and saves a desired visualization, the transformed data is used to resolve unknown concepts: these concepts are now available as known concepts to be used to create visualizations. 

\new{
\subsection{Implementation} Data Formulator is built as a React web application in Typescript; its backend is a Python server that runs on a Dv2-series CPU with 3.5 GiB RAM on Azure. Data Formulator's backend queries the OpenAI Codex API for concept derivation and runs the synthesis algorithm locally. Data Formulator's scalability to larger data relates to (1) the frontend's visualization rendering capability and (2) the backend's efficiency to execute data transformation scripts. To scale up Data Formulator for large datasets, we envision a sampling-based approach~\cite{moritz2019falcon}, where Data Formulator presents results on a representative subset of data to enhance interactivity and returns full results asynchronously.
}

\section{Evaluation: Chart Reproduction Study}\label{sec:user-study}

We conducted a chart reproduction study~\cite{ren2018reflecting} to gather feedback on the new concept binding approach that employs an AI agent, and to evaluate the usability of Data Formulator. 

\subsection{Study Design}
\noindent \textbf{Participants.} We recruited 10 participants (3 female, 7 male) from a large technology company. All participants had experience creating (simple) charts and identified themselves as a person with normal or corrected-to-normal vision, without color vision deficiency. Six participants are data scientists, two are applied scientists, and the remaining two are data \& applied scientists, and they are all located in the United States. Four participants are in their 30's, three are in 20's, and one participant is in each of the 40's, 50's, and 18-19 age group. They had varying levels of self-identified expertise in terms of chart authoring, computer programming, and experience with LLMs. 

\bpstart{Tasks and Datasets} We prepared six chart reproduction tasks with two datasets (3 tasks for each dataset): daily COVID-19 cases from Jan 21, 2020 to Feb 28, 2023 (3 columns; 1,134 rows) for the first task set (Tasks 1-3) and daily temperatures in 2020 for Seattle and Atlanta (4 columns; 732 rows; \cref{fig:sea-atl-temp-simple}) for the second set (Tasks 4-6). In both task sets, each subsequent task is built upon the previous one. One task (Task 4) required building two new concepts for reshaping and the other five tasks required the use of derived concepts. 
We also prepared three tutorial tasks, using students' exam scores dataset (5 columns; 1,000 rows): in addition to the scores for three subjects (math, reading, and writing), the data table included a student's id and major. The first tutorial task was about creating a chart with known/available concepts, while the second and third tutorial tasks were for creating charts using derived concepts and unknown concepts, respectively. Finally, we produced two practice tasks (one for reshaping and another for derivation). For these, the exam scores dataset was transformed into a long format, including math and reading scores under the subject and score column, resulting in 4 columns and 2,000 rows.

\bpstart{Setup and Procedure} We conducted sessions remotely via the Microsoft Teams. Each session consisted of four segments: (1) a brief explanation of the study goals and procedure, (2) training with tutorial and practice, (3) chart reproduction tasks, and (4) debrief. 

The training segment started with a quick introduction of Data Formulator's basic interactions using a simple task that does not require data transformation. Then, with their screen shared and recorded with audio, participants went through a tutorial and created three visualizations following step-by-step instructions provided in slides. They next created two visualizations on their own as practice.  
After an optional break, the participants performed six reproduction tasks using the two datasets mentioned above. Each task included a description (e.g., ``Create a Scatter Plot to compare Atlanta Temperature against Seattle Temperature.''), the labels for axes and color legend (if necessary), and an image of the target visualization. (Study materials are included in the supplemental material.) We encouraged the participants to think aloud, describing their strategies, whether any feature of Data Formulator works or makes sense, if the system behaves as they expect, etc. 
We recorded if the participants required a hint (and which hint) and how long it took for them to complete the task. The recorded completion time is not intended to indicate performance, as we wanted to gain insights about our approach using the think aloud method. Instead, we wanted to see if and how the participants completed, faltered, or recovered for each task, within a reasonable amount of time.
The session ended with a debriefing after the participants filled out a questionnaire with 5 questions about their experience with Data Formulator. The entire session took about two hours to complete, while the training segment took about an hour. We compensated each participant with a \$100 Amazon Gift card.

\subsection{Results}

After an hour-long tutorial and practice session, most participants could use Data Formulator to create different types of charts that involve advanced data transformations. Furthermore, they were generally positive about their experience with Data Formulator in chart authoring.

\bpstart{Tasks Completion and Usability Issues}
Participants completed all tasks on average within 20 minutes, with a deviation of about four and a half minutes. 
\Cref{table:taskcompletion} shows the average and standard deviation of task completion time in minutes, along with the total number of hints provided for each chart reproduction task (for all 10 participants). The participants spent most of their time (on average less than five minutes) on Task 6 because it was not trivial to inspect the code to generate 7-day moving average.
For Tasks 5 and 6, we had to give one hint (to two different participants) to guide them to use a different type of concept (they needed to derive a concept but initially tried to build a concept). There were a few cases that we had to provide a hint to a single participant: how to select multiple sources for derivation (Task 4), what are the correct source concepts for derivation (Tasks 2 \& 5), and the example values should be from the original table (Task 4). We had to provide the highest number of hints for Task 1. This was because when participants derived the year from the date value, its data type was set to number and the participants did not know or remember how to change its data type to string. (As detailed below, some participants tried to fix it by providing a different natural language prompt). 

\new{For derived concepts, once the participants identified the correct interaction approach and input fields, they are able to describe and refine the transformation in natural language to solve the tasks. We recorded all participants' prompts (see supplementary material). On average, participants made 1.62 prompt attempts per derived concept, and the length of those prompts averaged 7.28 words. The system generated an average of 1.94 candidates per prompt attempt.}

\begin{table}[h]
\small
\caption{The average and standard deviation of task time (in minutes) and the total number of hints provided for chart reproduction tasks.}
\centering
\renewcommand{\arraystretch}{0.7}
\begin{tabularx}{.44\textwidth}{c|c|c|c}
    \toprule
    Task & Average Time & Standard Deviation & Total Number of Hints \\ 
    \midrule
    Task 1 & 2:21 & 0:45 & 7 \\
    Task 2 & 3:19 & 2:09 & 2 \\
    Task 3 & 3:45 & 1:33 & 2 \\
    \midrule
    Task 4 & 2:43 & 1:33 & 2 \\
    Task 5 & 2:22 & 1:55 & 3 \\
    Task 6 & 4:29 & 1:39 & 2 \\
    \bottomrule
\end{tabularx}
\label{table:taskcompletion}
\vspace{-4mm}
\end{table}


Participants rated Data Formulator on five criteria using a 5-point Likert scale (5 being the most positive) as follows: easy to learn (\textit{M} = 3.90, \textit{SD} = 0.88), easier than other tools to transform data (\textit{M} = 3.80, \textit{SD} = 1.23), AI-agent's usefulness (\textit{M} = 4.4, \textit{SD} = 0.70), helpful to verify generated data (\textit{M} = 4.1, \textit{SD} = 0.74), and the trustworthiness of generated data (\textit{M} = 4.7, \textit{SD} = 0.48).

Participants provide feedback to improve the user interface. Four participants expected a way to multi-select on concept cards and click ``derive'' for deriving a concept from multiple existing ones. The current method of clicking ``derive'' on one concept and then multi-selecting is not intuitive. Two other participants expected the AI to select or identify which concepts to derive from based on their prompts. A few participants expected to change data type using the prompt (e.g., ``year as a string'' when the year is extracted from date).
Five participants wanted the derived examples table to show more values, or unique derived values. 
Reshaping data was at times a point of confusion: two participants found it difficult to understand how the AI formulated candidate datasets, while two others did not intuit or remember the task sequence to formulate data for unknown concepts.
When required to reshape data, three participants entered plausible, but not exact values in the example table during the training: they misunderstood the rigid connection to the original dataset.
To strengthen that connection participants recommended including additional columns (especially a column that is unique for a pivot transform) or to filter or highlight rows of the data table view that correspond to the values used in the example table. 
We also observed users' attempts to re-use a derived concept as a commutative function on other concepts: two participants tried to drag a derived concept and drop it on other concepts.

\bpstart{Overall Reaction and Experience} To understand participants' reaction to the new concept-drive approach employing an AI agent, we analyzed the debrief interview, during which participants stated something or confirmed an observation made by the experimenter. Using the transcription from the recorded sessions, one researcher applied an open coding method to surface all unique feedback, issues and ideas from the participants. 
He expanded the codes to generalize for semantically similar participant statements. While quantities of qualitative data does not provide a metric for importance, we counted how many participants mentioned each code, providing how frequently our participants had shared experiences or ideas with Data Formulator. 


Overall, participants were positive about their experience with Data Formulator. All 10 participants said that natural language prompts work well for generating data transforms and eight mentioned that AI is a helpful tool for the study tasks. Participants more frequently praised the derived concept than the unknown concept method for transforming data. Specifically, when it comes to verifying candidate derived concepts: all except one participant commented that displaying code was helpful and seven found the example derived values table to be useful. While only half of the participants commented that pivoting with unknown concepts is easier than with other tools, only three affirmed the example data table being helpful. 

Five participants mentioned that they were impressed by the power of the AI agent to generate data transforms. Five participants found having candidates (for both derived and formulated data) to be helpful because the candidates provided an opportunity to choose a correct answer, or at the least to select a promising direction to refine. Participants also explained that generating candidates increases trust in a collaborative experience. 
On the other hand, three participants mentioned they are reluctant to give much trust to the AI generative features of the tool.
\section{Related Work}
\label{sec:related-work}

Data Formulator builds on top of prior research in visualization authoring tools, data transformation tools, and code generation techniques.

\bpstart{Visualization Grammars and Tools} The grammar of graphics~\cite{DBLP:books/daglib/0024564} first introduces the representation of visualizations based on chart types and encodings of data columns to their visual channels. Many high-level grammars are designed to realize this idea. For example, ggplot2~\cite{wickham2009ggplot2} is a charting library in R based on visual encodings. Vega-Lite~\cite{satyanarayan2017vegalite} and its Python-wrapper Altair~\cite{vanderplas2018altair} extend the traditional grammar of graphics design with rules for layered and multi-view displays, as well as interactions, and Animated Vega-Lite~\cite{zong2022animated} further extends it to support animations. These grammars hide low-level implementation details and are concise. Therefore, they are generally preferred for the rapid creation of visualization in exploratory settings over toolkits and libraries like Protovis~\cite{bostock2009protovis}, Atlas~\cite{liu2021atlas}, and D3~\cite{bostock2011d3} that are designed for more expressive and novel visualization authoring. High-level grammars inspire interactive visualization tools like Tableau~\cite{stolte2002query}, Power BI, Lyra~\cite{satyanarayan2014lyra}, Charticulator~\cite{ren2019charticulator}, and Data Illustrator~\cite{liu2018data}. These tools adopt a shelf-configuration design: authors map data columns to visual encoding ``shelves'' often using the drag-and-drop interaction, and enerate specifications in high-level grammars to render visualizations. 
These grammars and tools require that the input data is in a tidy format, where all variables to be visualized are columns of input data. Because this means authors often need to transform the data first to create any visualizations, Satyanarayan et al. recognized the automatic inferring or suggestions of appropriate transformations when necessary, as an important research problem~\cite{satyanarayan2019critical}. 

To reduce authors' efforts, visualization by demonstration~\cite{zong2020lyra,saket2016visualization,shen2022galvis} and by example~\cite{wang2021falx} tools are introduced. Lyra 2~\cite{zong2020lyra} generates interaction rules after authors perform an interaction on the visualization. VbD~\cite{saket2016visualization} lets users demonstrate transformations between different types of visualizations to produce new specifications. Although these approaches reduce the chart specification efforts, they require tidy input data. Falx~\cite{wang2021falx}, on the other hand, addresses the data transformation challenge with a visualization-by-example design. Falx lets authors specify visualizations via low-level example mappings from data points to primitive chart elements. However, Falx does not support derivation types of transformation because of its underlying programming-by-example algorithm limitations; its requirement to focus on low-level elements also introduces a challenging paradigm shift for users who are more familiar with tools that focus on high-level specifications~\cite{satyanarayan2017vegalite,stolte2002query}.

\new{Natural language interfaces~\cite{chen2022type,luo2021natural,poesia2022synchromesh,DBLP:conf/uist/GaoDALK15,DBLP:conf/chi/KimHA20,DBLP:conf/apvis/0004HJY21} enhance users' ability to author and reason about visualizations.  NCNet~\cite{luo2021natural} uses a Seq-to-Seq model to translate chart description texts into Vega-Lite specs. VisQA~\cite{DBLP:conf/chi/KimHA20} is a pipeline that leverages semantic parsing techniques~\cite{DBLP:conf/acl/PasupatL15} to provide atomic data-related answers based on its visualizations. NL4DV~\cite{narechania2020nl4dv} and Advisor~\cite{DBLP:conf/apvis/0004HJY21} generate visualizations based on user questions. To manage ambiguity in natural language inputs~\cite{srinivasan2021collecting}, DataTone~\cite{DBLP:conf/uist/GaoDALK15} ranks solutions based on user preference history, and Pumice~\cite{DBLP:conf/uist/LiRJSMM19pumice} introduces a multi-modal approach that leverages examples to refine the initial ambiguous specification. Data Formulator's concept derivation interface is based on natural language. Data Formulator benefits from large language models' expressiveness~\cite{chen2021evaluating}, and manages ambiguity by restricting the target function type to columns-to-column mapping functions (as opposed to arbitrary data transformation scripts). In the future, more powerful language models can be plugged into Data Formulator to improve code generation quality.}

Data Formulator adopts the shelf-configuration approach like Tableau and Power BI, but it supports encoding from \emph{data concepts} to visual channels to address the data transformation burden. Because Data Formulator can automatically transform the input data based on the concepts used in the visualization specification, authors do not need to manually transform data. Furthermore, because Data Formulator's Chart Builder resembles tools like Power BI and Tableau, it lets the authors focus on high-level designs. Data Formulator's multi-modal interaction approach supports both derivation and reshaping tables. While Data Formulator currently focuses on standard visualization supported by Vega-Lite, its AI-powered concept-driven approach can also work with expressive and creative visualization design tools like StructGraph~\cite{tsandilas2020structgraphics} and Data Illustrator~\cite{liu2018data} to automate data transformations.

\bpstart{Data Transformation Tools} Libraries and tools like tidyverse~\cite{wickham2019tidyverse}, pandas~\cite{the_pandas_development_team_2023_7741580}, Potter's Wheel~\cite{raman2001potter}, Wrangler~\cite{kandel2011wrangler}, Tableau Prep, and Power Query are developed to support data transformation. They introduce operators to reshape, compute, and manipulate tabular data needed in data analysis. Automated data transformation tools, including programming-by-example tools\cite{polozov2015flashmeta,wang2017synthesizing,jin2017foofah} and initiative tools~\cite{jin2020auto,yan2020auto,beth2020mage,kandel2011wrangler}, are developed to reduce authors' specification effort. Data Formulator tailors key transformation operators from the tidyverse library (reshaping and derivation) for visualization authoring. Because the desired data shape changes with visualization goals, even with these tools, authors still need the knowledge and effort to first identify the desired data shape, and then switch tools to transform the data. Data Formulator bridges visual encoding and data transformation with data concepts to reduce this overhead.

\bpstart{Code Generation} Code generation models~\cite{chen2021evaluating,chowdhery2022palm,fried2022incoder} and program synthesis techniques~\cite{gulwani2017program,wang2017synthesizing,chaudhuri2021neurosymbolic,zhang2021interpretable} enable users to complete tasks without programming by using easier specifications, including natural language, examples, and demonstrations. Code generation models like Codex~\cite{chen2021evaluating}, PaLM~\cite{chowdhery2022palm}, and InCoder~\cite{fried2022incoder} are transformer-based causal language models (commonly referred to as LLMs) that complete texts from natural language prompts. These LLMs can generate expressive programs to solve competitive programming~\cite{li2022competition,hendrycks2021measuring}, data science~\cite{lai2022ds}, and software engineering tasks~\cite{barke2022grounded} from high-level descriptions. Programming-by-example~\cite{wang2021falx} and programming-by-demonstration~\cite{barman2016ringer,pu2022semanticon} tools can synthesize programs based on users' output examples or demonstrations that illustrate the computation process. Natural language approaches are highly expressive, but some tasks can be challenging to phrase. On the other hand, while programming-by-example techniques are precise, they are less expressive and do not scale to large programs as they require well-defined program spaces. Therefore, Data Formulator adopts a mixed-modality approach to solve the data transformation task. It leverages the Codex model~\cite{chen2021evaluating} for concept derivations and the example-based synthesis algorithm~\cite{wang2019visualization} for reshaping, which takes advantage of both approaches to reduce authors' specification overhead.

Because code generation techniques generalize programs from incomplete user specifications, generated programs are inherently ambiguous, and thus require disambiguation to identify a correct solution among candidates. Prior work proposes techniques to visualize the search process~\cite{zhang2021interpretable}, visualize code candidates~\cite{wang2021falx,xiong1912revealing}, and present distinguishing examples for authors to inspect~\cite{ji2020question}. Data Formulator provides feedback to the authors by presenting the generated code together with its execution results for them to inspect, select, and edit.
\section{Discussion and Future Work}\label{sec:discussion}



\noindent \textbf{Unified Interaction with Multiple Modalities.} Data Formulator employs two different modalities for authors to specify different types of data transformation: natural language for concept derivation and examples for table reshaping (\cref{sec:design}). This design combines strengths of both modalities so that the authors can better communicate their intent with the AI agent, and the AI agent can provide precise solutions from a more expressive program space. However, choosing the right input modality when creating a new concept can be challenging for inexperienced authors. To address this challenge, we envision a stratified approach where the authors just initiate the interaction in natural language, and the AI agent will decide whether to ask the authors, for example relations for clarification or to directly generate derivation codes. This design will shift the effort of deciding which approach to start with from the authors to the AI agent, and ``by-example'' specification will become a followup interaction step to help the authors clarify their intent. We envision this mixed-initiative unified interaction will further reduce the authors' efforts in visualization authoring.

\bpstart{Conversational Visualization Authoring with AI Agents} Conversational AI agents~\cite{ouyang2022training} have the strength of leveraging the interaction contexts to better interpret user intent. They also provide opportunities for users to refine the intent when the task is complex or ambiguous. However, conversation with only natural language is often insufficient for visualization authoring because (1) it does not provide users with precise control over the authoring details (e.g., exploring different encoding options, changing design styles) and (2) the results can be challenging to inspect and verify without concrete artifacts (e.g., programs, transformed data). It would be useful to research how conversational AI can be integrated with Data Formulator's concept-driven approach to improve the overall visualization experiences. First, with a conversational AI agent, the authors can incrementally specify and refine their intent for tasks that are difficult to solve in one shot. Second, a conversational agent complements Data Formulator by helping the authors explore and configure chart options. Because Data Formulator focuses on data transformation, it does not expose many chart options (e.g., axis labeling, legend style, visual mark styles) in its interface. A conversational AI agent can help the authors access and control these options without overwhelming them with complex menus. For example, when the authors describe chart styles they would like to change, Data Formulator can apply the options directly or dynamically generate editing panels for them to control. We envision the effective combination of conversational AI experiences, and the Data Formulator approach will let the authors confidently specify richer designs with less effort. 

\bpstart{Concept-driven Visual Data Exploration} Visual data exploration tools~\cite{wongsuphasawat2015voyager,Wongsuphasawat2017voyager2,moritz2018formalizing,lee2021lux,lee2021deconstructing} help data scientists understand data and gain meaningful insights in their analysis process. 
These tools support a rich visual visualization space, yet still require datasets to be in the appropriate shape and schema. While Data Formulator is designed for visualization authoring, its concept-driven approach can be used in visual data exploration to expand the design space. Beyond the current concept-driven features of Data Formulator, the AI agent could be enhanced to recommend data concepts of interest based on the data context or author interaction history. Building on this idea, the tool could recommend charts based on all potentially relevant data concepts. This expansive leap could overcome one of the limitations of chart recommendation systems: by enabling the authors to view charts beyond their input data columns without additional user intervention. 

\bpstart{Study Limitations}
While our participants had varying levels of expertise in chart authoring, computer programming, and experience with LLMs, many of them had considerable knowledge about data transformation methodology and programming. It would be useful to investigate if and how people with limited expertise could learn and use Data Formulator.
The main goal of Data Formulator was to reduce manual data transformation in visualization authoring efforts. As such, in our study, we focused on derivation and reshaping types of data transformations with simple datasets. While they are key types of transformation and our tasks covered multiple styles of derivations, the transformations we studied are by no means comprehensive. 
It would be valuable to evaluate the broader combinations and complexities of data transformations.
Our study adopted a chart reproduction study~\cite{ren2018reflecting}, which is commonly used for evaluating chart authoring systems (e.g., ~\cite{liu2018data,ren2019charticulator,satyanarayan2014lyra}). Therefore, our study shares its inherent limitations: because we prepared datasets and tasks, and provided target visualizations as a reference, we do not know if and how people would use Data Formulator to create visualizations with their own data.  
\section{Conclusion} This paper introduces Data Formulator, a concept-driven visualization authoring tool that leverages an AI agent to address the data transformation challenge in visualization authoring. With Data Formulator, authors work with the AI agent to create new data concepts and then map data concepts to visual channels to specify visualizations. Data Formulator then automatically transforms the input data and instantiates the visualizations. Throughout the authoring process, Data Formulator provides feedback to the authors to inspect and refine the generated code to promote confidence. As discovered in the chart reproduction study, participants can learn and use Data Formulator to create visualizations that require advanced data transformations. In the future, the concept-driven visualization approach can potentially benefit the design of new visual data exploration tools and expressive visualization authoring tools to overcome the data transformation barrier.


\appendix
\section{Supplemental Material} 

We include three zip files in the supplemental material: (1) a 6-minute video that walks through user experiences of creating visualizations about Seattle and Atlanta temperatures, described in \cref{sec:user-experience} with Data Formulator, (2) a set of short videos that demonstrate additional Data Formulator scenarios, and (3) our user study materials including: study script, tutorials, study tasks, and prompts created by participants.

\bibliographystyle{abbrv-doi-hyperref}

\bibliography{dataformulator}


\end{document}